\documentclass [aps,prb,preprint,showpacs]{revtex4}
\usepackage{graphicx}
\usepackage{amsmath}
\usepackage{braket}
\usepackage{xcolor}

\begin{document}

\title[High-efficiency triple-resonant inelastic light scattering in planar optomagnonic cavities]{High-efficiency triple-resonant inelastic light scattering in planar optomagnonic cavities}

\author{Petros Andreas Pantazopoulos}
\email{pepantaz@phys.uoa.gr}
\author{Kosmas L. Tsakmakidis}
\author{Evangelos Almpanis}
\author{Grigorios P. Zouros}
\author{Nikolaos Stefanou}
\affiliation{Section of Solid State Physics,
	National and Kapodistrian University of Athens, \\
	Panepistimioupolis, GR-157~84 Athens, Greece}
\date{\today}


\vspace{10pt}

\begin{abstract}
Optomagnonic cavities have recently been emerging as promising candidates for implementing coherent microwave-to-optical conversion, quantum memories and devices, and next generation quantum networks. 
A key challenge in the design of such cavities is the attainment of high efficiencies, which could, e.g., be exploited for efficient optical interfacing of superconducting qubits, as well as the practicality of the final designs, which ideally should be planar and amenable to on-chip integration. 
Here, on the basis of a novel time Floquet scattering-matrix approach, we report on the design and optimization of a planar, multilayer optomagnonic cavity, incorporating a Ce:YIG thin film, magnetized in-plane, operating in the triple-resonant inelastic light scattering regime.
This architecture allows for conversion efficiencies of about 5\%, under realistic conditions, which is orders of magnitude higher than alternative designs. 
Our results suggest a viable way forward for realizing practical information inter-conversion between microwave photons and optical photons, mediated by magnons, with efficiencies intrinsically greater than those achieved in optomechanics and alternative related technologies, as well as a platform for fundamental studies of classical and quantum dynamics in magnetic solids, and implementation of futuristic quantum devices.
\end{abstract}

%

%
\maketitle
%
%

\section{Introduction}
Optomagnonic cavities are judiciously designed dielectric structures that include magnetic materials capable of simultaneously confining light and spin waves in the same region of space.
This confinement leads, under certain conditions, to strong enhancement of the inherently weak interaction between the two fields, which allows for an efficient microwave-to-optical transduction, enabling, e.g., optical interfacing of superconducting qubits~\cite{tabuchi2015,lachance2017}.

\begin{figure}
\centering
\includegraphics[width=1\linewidth]{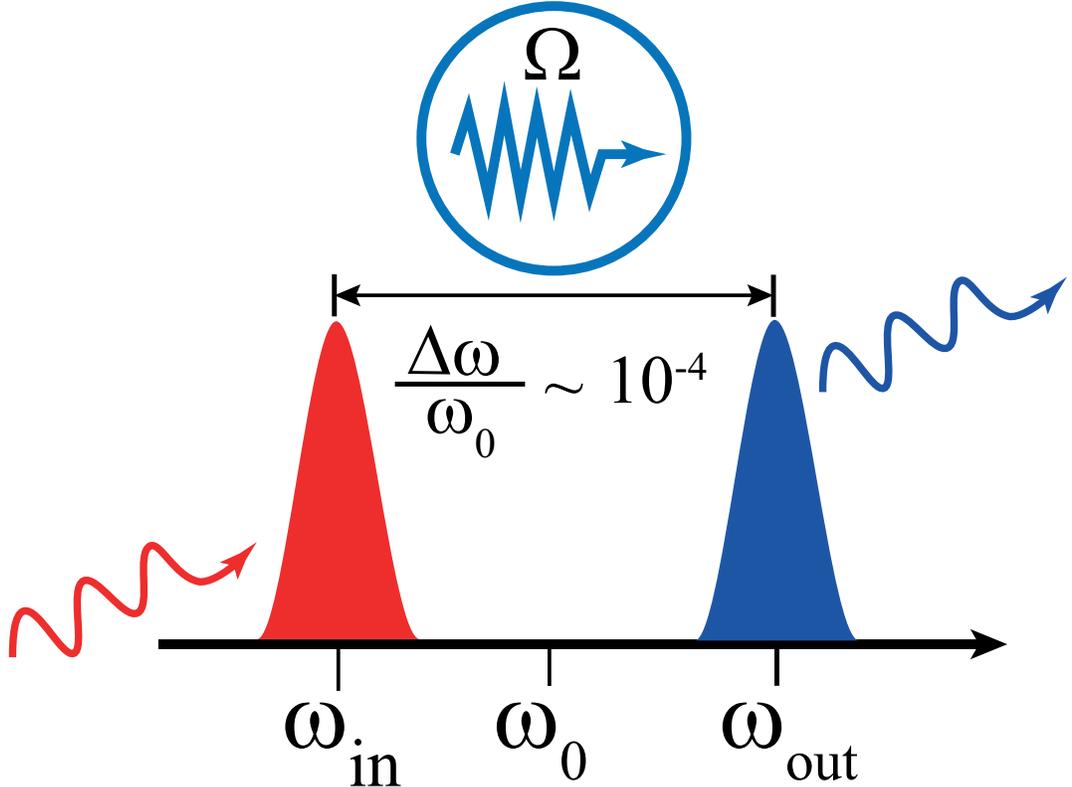}
\caption{Schematic of inelastic light scattering through magnon absorption in an optomagnonic cavity. The frequency of a cavity magnon, $\Omega$, matches a photon transition between two resonant optical modes:
$\Omega=\Delta\omega\equiv\omega_\mathrm{out}-\omega_\mathrm{in}$ (triple-resonance condition). }
\label{fig1}
\end{figure}

The optomagnonic interaction is expected to be larger when the so-called triple-resonance condition is met, i.e., when the frequency of a cavity magnon matches a photon transition between two resonant modes.
This implies that the cavity must support two well-resolved optical resonances (in the hundred terahertz range) separated by a few gigahertz, which requires quality factors at least of the order of $10^5$, as schematically depicted in figure~\ref{fig1}.

A (sub)millimeter-sized sphere, made of a low-loss dielectric magnetic material, constitutes a simple realization of an optomagnonic cavity. The sphere supports densely spaced long-lifetime optical whispering gallery modes~\cite{osada2016,zhang2016,haigh2016,kusminskiy2016,sharma2017}, and infrared incident light evanescently coupled to these modes can be scattered by a uniformly precessing (so-called Kittel) spin wave to a neighbouring optical whispering gallery mode.
In the prospect of achieving smaller modal volumes and larger spatial overlap between the interacting fields, higher-order magnetostatic modes~\cite{haigh2018,osada2018A,osada2018B}, magnetically split optical Mie resonances in small spheres~\cite{almpanis2018}, as well as particles of different shapes~\cite{graf2018} have been proposed.
However, these proposals currently face appreciable challenges in the fabrication of high-quality particles and/or the efficient excitation of the spin waves.

A promising alternative design of optomagnonic cavities is based on \textit{planar} geometries, which can exhibit even stronger magnon-to-photon conversion efficiencies~\cite{kostylev2019}, while at the same time allowing integration into a hybrid opto-microwave chip using modern nanofabrication methods.
To this end, optomagnonic cavities formed in a magnetic dielectric film bounded by two mirrors~\cite{liu2016,pantazopoulos2017,pantazopoulos2019B}, or in a defect layer in a dual photonic-magnonic periodic layered structure~\cite{pantazopoulos2019A}, have also been investigated. 
However, the studies reported so far refer to the Faraday configuration, with out-of-plane magnetized films, where it is challenging to obtain two optical resonances in the required close proximity to each other. 

In this work we show that, by using in-plane magnetized films in the so-called Voigt configuration, we can overcome the afore-described shortcomings of previous schemes and design efficient optomagnonic cavities operating in the triple-resonance regime. In section~\ref{sc:design} we describe our statically magnetized structure and discuss its optical response. 
In section~\ref{sc:theory} we summarize our recently developed fully dynamic time Floquet method for layered optomagnonic structures~\cite{pantazopoulos2019B} and in section~\ref{sc:results} we present details of our attained numerical results. The last section concludes the article.

\section{Structure Design}\label{sc:design}
\begin{figure}
\centering
\includegraphics[width=\linewidth]{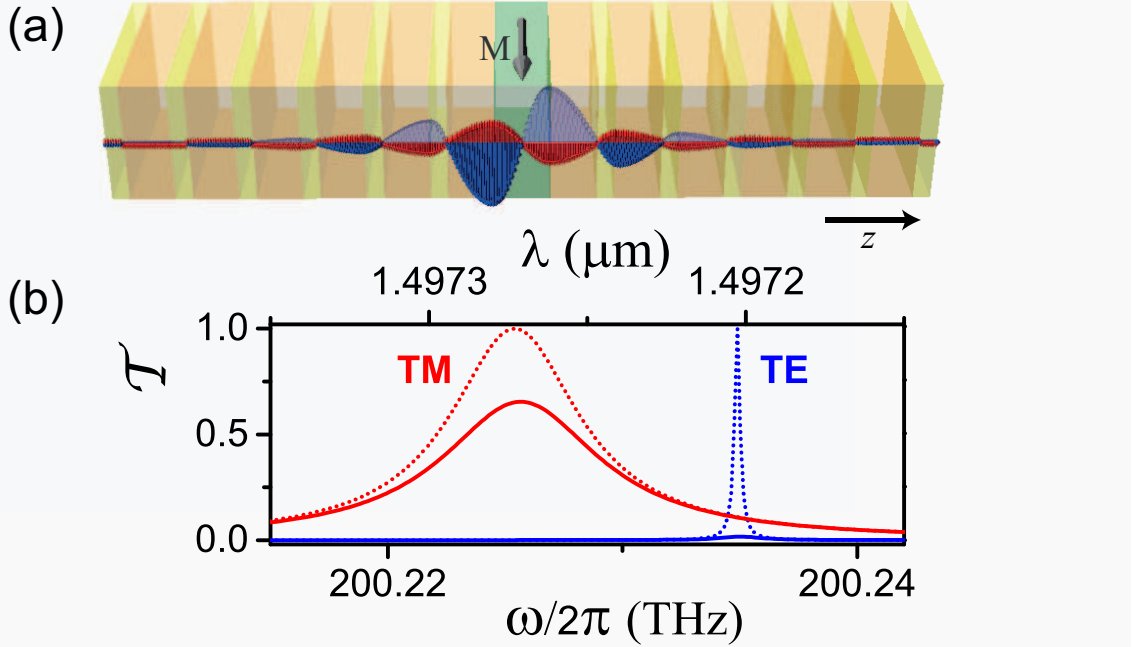}
\caption{ (a) Schematic view of the optomagnonic cavity under study. 
It is formed by a 350-nm-thick Ce:YIG film magnetized in-plane (along the $x$ direction), bounded symmetrically by two Bragg mirrors, each consisting of 6 periods of alternating $\mathrm{SiO}_2$ and Si layers of thickness 290~nm and 110~nm, respectively, grown along the $z$ direction. 
For light incident with $q_y=3~\mu m^{-1}$, the cavity supports two localized resonant modes, one of TM and the others of TE polarization, manifested in the corresponding transmission spectrum shown in (b), with the dotted and solid curves referring to the lossless and lossy structure, respectively. A snapshot of the associated electric field profiles along the $z$ direction in the lossless case is illustrated in (a).}
\label{fig2}
\end{figure}

We propose a simple design of planar optomagnonic cavity, simultaneously confining light and spin waves in the same subwavelength region of space.
It consists of an iron garnet thin film bounded symmetrically by two-loss, dielectric Bragg mirrors, in air, as schematically illustrated in~figure~\ref{fig2}(a).

Iron garnets are ferrimagnetic materials exhibiting important functionalities for bulk and thin-film device applications that require magnetic insulators, owing to their unique physical properties such as high optical transparency in a wide range of wavelengths, high Curie temperature, ultra-low spin-wave damping, and strong magneto-optical coupling~\cite{zvezdin_book}.
In our work, we consider cerium-substituted yttrium iron garnet (Ce:YIG) which, at the telecom wavelength of $1.5~\mu\mathrm{m}$, has a relative electric permittivity $\epsilon=5.10+i4\times10^{-4}$ and a Faraday coefficient $f=-0.01$~\cite{onbasli2016}, while its relative magnetic permeability equals unity. 
The Ce:YIG film extends from $-d/2$ to $d/2$ and is magnetically saturated to $M_0$ by an in-plane bias magnetic field $H_0$ oriented, say, along the $x$ direction. 
Therefore, the corresponding relative electric permittivity tensor, neglecting the small Cotton-Mouton contributions, is of the form~\cite{stancil_book}
\begin{equation}\label{eq:epsilon}
   {\boldsymbol{\epsilon}}=
   \left(\begin{array}{ccc}
     \epsilon & 0&0\\
     0&\epsilon&if\\
     0&-if&\epsilon
   \end{array}\right)\;.
\end{equation}

We consider the Voigt geometry with light propagating in the $y$-$z$ plane. The structure in this geometry, with the magnetic field parallel to the surface and also perpendicular to the propagation direction, remains invariant under reflection with respect to the plane of incidence.
Consequently, contrary to the Faraday configuration studied in our previous work~\cite{pantazopoulos2017,pantazopoulos2019B,pantazopoulos2019A}, the transverse magnetic (TM) and transverse electric (TE) polarization modes, i.e., modes with the electric field oscillating in and normal to the plane of incidence, respectively, are eigenmodes of the system.
Interestingly, in the chosen geometry, the magnetic film behaves as isotropic, with permittivity $\epsilon-f^2/\epsilon$ and $\epsilon$ for TM- and TE-polarized waves, respectively.
In other words, only TM-polarized light is affected by the (magnetic) polarization field. 

Each Bragg mirror consists of an alternate sequence of six $\mathrm{SiO_2}$ and six Si quarter-wave layers, i.e., $d_\mathrm{m}\sqrt{(2\pi n_\mathrm{m}/\lambda)^2-q_y^2}=\pi/2$, where $d_\mathrm{m}$ (m: Si$\mathrm{O_2}$ or m: Si) is the layer thickness and $n_\mathrm{m}$ the corresponding refractive index ($n_\mathrm{SiO_2}=1.47$ and $n_\mathrm{Si}=3.5$) at the operation wavelength $\lambda\approx1.5~\mu\mathrm{m}$~\cite{pierce1972,gao2012}.
Due to translation invariance parallel to the $x$-$y$ plane, the in-plane component of the wave vector, $q_y=2\pi\sin\theta/\lambda$, where $\theta$ is the angle of incidence, remains constant. 
Taking, for instance, $q_y=3~\mu\mathrm{m}^{-1}$, which corresponds to an angle of incidence of about $45^\mathrm{o}$, we obtain $d_\mathrm{SiO_2}=290$~nm and $d_\mathrm{Si}=110$~nm.
Accordingly, we choose a thickness $d=350$~nm for the Ce:YIG film to satisfy the half-wave condition that corresponds to transmission maxima.

This design provides two (one TM and one TE) high-quality-factor resonances within the lowest Bragg gap, at a wavelength of about 1.5~$\mu$m, separated by a frequency difference $\Delta f=9.5$~GHz that matches the frequency of magnetostatic spin waves~\cite{stancil_book,zvezdin_book}.
These resonant modes are strongly localized in the region of the Ce:YIG film, which can be considered as a defect in the periodic stacking sequence of the Bragg mirrors. 
Absorption losses reduce the transmittance peak.
In particular, the long-lifetime TE resonance is strongly suppressed in the presence of dissipative losses, as shown by the solid line blue line in~figure~\ref{fig2}(b).

It should be pointed out that the position and width of the optical resonances can be tailored at will by appropriate selection of the materials, and by properly adjusting the geometric parameters of the structure and the angle of incidence.

\section{Theory for Layered Optomagnonic Structures}\label{sc:theory}
The magnetic Ce:YIG film supports magnetostatic spin waves where the magnetization precesses in-phase, elliptically, throughout the film (uniform precession mode) with angular frequency 
$\Omega=\sqrt{\Omega_\mathrm{H}(\Omega_\mathrm{H}+\Omega_\mathrm{M})}$,
where $\Omega_\mathrm{H}=\gamma\mu_0H_0$ and $\Omega_\mathrm{M}=\gamma\mu_0M_0$, $\gamma$ being the gyromagnetic ratio and $\mu_0$ the magnetic permeability of vacuum~\cite{stancil_book}.
The corresponding magnetization field profile is given by 
\begin{equation}\label{eq:sw_field}
\mathbf{M}(\mathbf{r},t)/M_0=\widehat{\mathbf{x}}+\eta A_{y} \mathrm{sin}(\Omega t)\widehat{\mathbf{y}}+ \eta A_{z} \mathrm{cos}(\Omega t)\widehat{\mathbf{z}}\;,
\end{equation}
where $
A_{y}=\sqrt{(\Omega_\mathrm{H}+\Omega_\mathrm{M})/(2\Omega_\mathrm{H}+\Omega_\mathrm{M})}$, $
A_{z}=\sqrt{\Omega_\mathrm{H}/(2\Omega_\mathrm{H}+\Omega_\mathrm{M})}
$,
and $\eta$ is an amplitude factor that defines the magnetization precession angle.

Under the action of the spin wave, the magnetic film and, consequently, the entire structure can be looked upon as a periodically driven system because the magnetization field, given by Eq.~\eqref{eq:sw_field}, induces a temporal perturbation~\cite{pantazopoulos2017} 
\begin{equation}\label{eq:depsilon_decomp}
   \delta{\boldsymbol{\epsilon}}(t)=\frac{1}{2}\big[\delta{\boldsymbol{\epsilon}}\exp{(-i\Omega t)}+\delta{\boldsymbol{\epsilon}}^\dagger\exp{(i\Omega t)}\big]\;
\end{equation}
in the permittivity tensor of the statically magnetized material, where
\begin{equation}\label{eq:depsilon_zfunc}
   \delta{\boldsymbol{\epsilon}}=f\eta
   \left(\begin{array}{ccc}
     0 & iA_{z}&A_{y}\\
     -iA_{z}&0&0\\
     -A_{y}&0&0
   \end{array}\right)\;.
\end{equation}

The solutions of the underlying Maxwell equations are Floquet modes $\mathbf{F}(\mathbf{r},t)=\mathrm{Re}\{\boldsymbol{\mathcal{F}}(\mathbf{r},t)\exp(-i\omega t)\}$, with $\boldsymbol{\mathcal{F}}(\mathbf{r},t+T)=\boldsymbol{\mathcal{F}}(\mathbf{r},t)$, $T=2\pi/\Omega$, where by $\mathbf{F}$ we denote electric field, electric displacement, magnetic field, and magnetic induction, while $\omega$ is the Floquet quasi-frequency, similarly to the Floquet quasi-momentum (or else the Bloch wave vector) when there is spatial periodicity~\cite{comphy1,comphy2}.
Seeking Floquet modes in the form of plane waves with given $q_y$ and expanding all time-periodic quantities into truncated Fourier series in the basis of complex exponential functions $\exp(in\Omega t)$, $n=-N,-N+1,\ldots,N$, leads to an eigenvalue-eigenvector equation, which has $4(2N+1)$ physically acceptable solutions~\cite{pantazopoulos2019B}.
We characterize them by the following indices: $s=+(-)$ that denotes waves propagating or decaying in the positive (negative) $z$ direction, $p=1,2$ that indicates the two eigen-polarizations, and $\nu=-N,-N+1,\cdots,N$  which labels the different eigenmodes. 
These eigenmodes are polychromatic waves, each composed of $2N+1$ monochromatic components of angular frequency $\omega-n\Omega$, $n=-N,-N+1,\ldots,N$~\cite{pantazopoulos2019B}.
We note that, in a static homogeneous medium, the corresponding eigenmodes of the electromagnetic (EM) field are monochromatic waves characterized by the indices $s,p$, and $n$.

\begin{figure}[t]
  \centering
  \includegraphics[width=1\linewidth]{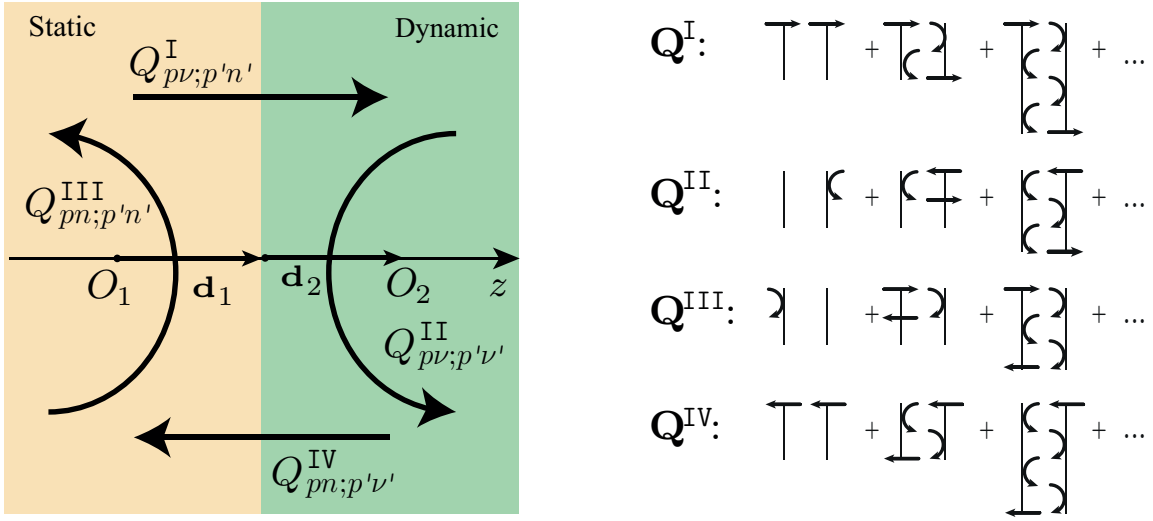}
  \caption{Left-hand diagram: Transmission and reflection matrices for a planar interface between a static and a dynamic medium, defined with respect to appropriate origins, $O_1$ and $O_2$, at distances
  $-\mathbf{d}_1$ and $\mathbf{d}_2$ from a center at the interface, respectively. Right-hand diagram: The transmission and reflection matrices of two consecutive interfaces are evaluated by summing up all relevant multiple-scattering processes.}
  \label{fig3}
\end{figure}

Scattering of an eigenmode occurs at an interface between two different homogeneous media. 
For such a planar interface between a static and a time-periodic medium, the relative complex amplitudes of the transmitted (reflected) waves, denoted by $Q^\mathrm{I}_{p\nu;p'n'}$ ($Q^\mathrm{III}_{pn;p'n'}$) for incidence in the forward direction or  $Q^\mathrm{IV}_{pn;p'\nu'}$ ($Q^\mathrm{II}_{p\nu;p'\nu'}$) for incidence in the backward direction in the configuration shown in~figure~\ref{fig3}, are obtained in the manner described in Ref.~\cite{pantazopoulos2019B}.
Primed indices refer to the incident wave.
For an interface between two static homogeneous media, the $Q$ matrices relate monochromatic waves and are diagonal in $n$, which reflects frequency conservation.
We note that, in order to evaluate the scattering properties of layered optomagnonic structures in a straightforward manner, the waves on each side of a given interface are expressed around different points, at a distance $-\mathbf{d}_1$ and $\mathbf{d}_2$ from the center of the interface (see figure~\ref{fig3}), so that all backward and forward propagating or evanescent waves in the region between two consecutive interfaces refer to the same (arbitrary) origin.
Of course, because of translation invariance parallel to the $x$-$y$ plane, the choice of the $x$-$y$ components of $\mathbf{d}_1$ and $\mathbf{d}_2$ are immaterial; thus, for simplicity, we choose $\mathbf{d}_1$ and $\mathbf{d}_2$ along the $z$ direction.

The transmission and reflection matrices of a pair of consecutive interfaces, $\mathrm{i}$ and $\mathrm{i}+1$, are obtained by properly combining those of the two interfaces so as to describe multiple scattering to any order. 
This leads to the following expressions after summing up the infinite geometric series involved, as schematically illustrated in figure~\ref{fig3}, i.e.,	

\begin{align}
	\nonumber &\mathbf{Q}^{\mathrm{I}}(\mathrm{i},\mathrm{i}+1) \!=\! \mathbf{Q}^{\mathrm{I}}(\mathrm{i}+1) [
	\mathbf{I} - \mathbf{Q}^{\mathrm{II}}(\mathrm{i})
	\mathbf{Q}^{\mathrm{III}}(\mathrm{i}+1)]^{-1} \mathbf{Q}^{\mathrm{I}}(\mathrm{i}) \\
	\nonumber &\mathbf{Q}^{\mathrm{II}}(\mathrm{i},\mathrm{i}+1) \!=\!
	\mathbf{Q}^{\mathrm{II}}(\mathrm{i}+1) \!+\! \mathbf{Q}^{\mathrm{I}}(\mathrm{i}+1)
	\mathbf{Q}^{\mathrm{II}}(\mathrm{i}) [ \mathbf{I}\! -\!
	\mathbf{Q}^{\mathrm{III}}(\mathrm{i}+1)
	\mathbf{Q}^{\mathrm{II}}(\mathrm{i})]^{-1} \mathbf{Q}^{\mathrm{IV}}(\mathrm{i}+1) \\
	\nonumber &\mathbf{Q}^{\mathrm{III}}(\mathrm{i},\mathrm{i}+1) \!=\!
	\mathbf{Q}^{\mathrm{III}}(\mathrm{i}) \!+\! \mathbf{Q}^{\mathrm{IV}}(\mathrm{i})
	\mathbf{Q}^{\mathrm{III}}(\mathrm{i}+1) [\mathbf{I} \!-\!
	\mathbf{Q}^{\mathrm{II}}(\mathrm{i})
	\mathbf{Q}^{\mathrm{III}}(\mathrm{i}+1)]^{-1} \mathbf{Q}^{\mathrm{I}}(\mathrm{i}) \\
	&\mathbf{Q}^{\mathrm{IV}}(\mathrm{i},\mathrm{i}+1) \!=\!
	\mathbf{Q}^{\mathrm{IV}}(\mathrm{i}) [ \mathbf{I} -
	\mathbf{Q}^{\mathrm{III}}(\mathrm{i}+1) \mathbf{Q}^{\mathrm{II}}(\mathrm{i})]^{-1}
	\mathbf{Q}^{\mathrm{IV}}(\mathrm{i}+1)\label{eq:q_multiple} \;.
\end{align}
It should be noted that the waves on the left (right) of the pair of interfaces are referred to an origin at a distance $-\mathbf{d}_{1}(\mathrm{i})$ [$\mathbf{d}_{2}(\mathrm{i}+1)$] from the center of the $\mathrm{i}$-th [$(\mathrm{i}+1)$-th] interface. 
We also recall that, though the choice of $\mathbf{d}_1$ and $\mathbf{d}_2$ associated to each interface is to a certain degree arbitrary, it must be such that ${d}_{2z}(\mathrm{i})+{d}_{1z}(\mathrm{i+1})$ equals the thickness of the layer between the i-th and (i+1)-th interfaces. 
It is obvious that one can repeat the above process to obtain the transmission and reflection matrices $\mathbf{Q}$ of three consecutive interfaces, by combining those of the pair of the first interfaces with those of the third interface, and so on, by properly combining the ${Q}$ matrices of component units, one can obtain the ${Q}$ matrices of a slab which comprises any finite number of interfaces~\cite{comphy1,comphy2}.
This method applies to an arbitrary slab which comprises periodically time-varying layers, provided that all dynamic media have the same temporal periodicity.
It is then straightforward to calculate the transmittance, $\mathcal{T}$, and reflectance, $\mathcal{R}$, of the slab as the ratio of the transmitted and reflected, respectively, energy flux to the energy flux associated with the incident wave.
$\mathcal{T}$ and $\mathcal{R}$ are given by the sum of the corresponding quantities over all scattering channels ($p,n$): $\mathcal{T}=\sum_{p,n}\mathcal{T}_{pn}$ and $\mathcal{R}=\sum_{p,n}\mathcal{R}_{pn}$.
It is worthnoting that, because of the time variation of the permittivity tensor, the EM energy is not conserved even in the absence of dissipative (thermal) losses. 
In this case, $\mathcal{A}=1-\mathcal{T}-\mathcal{R}>0 \,(<0)$ means energy transfer from (to) the EM to (from) the spin-wave field. 

We close this section by pointing out a useful polarization selection rule, which can be readily derived in the linear-response approximation. To first order, the coupling strength associated to the photon-magnon scattering is proportional to the overlap integral $G=\bra{\mathrm{out}}\delta{\boldsymbol{\epsilon}}\ket{\mathrm{in}}$, where $\braket{\mathbf{r}t|{\mathrm{in}}}=\mathbf{E}^{\mathrm{in}}(z)\exp[i(\mathbf{q}_\parallel\cdot\mathbf{r}-\omega t)]$ and $\braket{{\mathrm{out}}|\mathbf{r}'t'}=\mathbf{E}^{\mathrm{out}\star}(z)\exp[-i(\mathbf{q}_\parallel'\cdot\mathbf{r}-\omega' t)]$ denote appropriate incoming and outgoing monochromatic time-harmonic waves in the static magnetic layered structure.
Using Eq.~\eqref{eq:depsilon_decomp} we obtain
\begin{equation}
\label{eq:coupling}
G=4\pi^3f\eta\delta(\mathbf{q}_\parallel-\mathbf{q}_\parallel')\big[\delta(\omega-\omega'-\Omega)g_{-}+\delta(\omega-\omega'+\Omega)g_{+}\big]
\end{equation}
where $   g_{\pm}=\displaystyle\int dz \mathbf{u}_{\pm}\cdot\big[\mathbf{E}^{\mathrm{out}\star}(z)\times\mathbf{E}^{\mathrm{in}}(z) \big] $, with $\mathbf{u}_{\pm}=\mp A_y\widehat{\mathbf{y}}+iA_z\widehat{\mathbf{z}}$.
The delta functions in Eq.~\eqref{eq:coupling} express conservation of in-plane momentum and energy in inelastic light scattering processes that involve emission and absorption of one magnon by a photon, as expected in the linear regime.
Obviously, the amplitude of transition between two optical eigenmodes of the same polarization, TM or TE, is identically zero because the corresponding eigenvectors are real.
In other words, one-magnon processes change the linear polarization state of a photon.
\section{Results and Discussion}\label{sc:results}
We now assume continuous excitation of a uniform-precession spin-wave mode in the magnetic film, with a relative amplitude $\eta=0.06$, which induces a periodic time variation in the corresponding electric permittivity tensor, given by Eq.~\eqref{eq:depsilon_decomp}.
The optomagnonic structure is illuminated from the left by TM-polarized light with $q_y=3~\mu\mathrm{m}^{-1}$ at the corresponding resonance frequency, which corresponds to an angle of incidence of about $45^\mathrm{o}$.
The dynamic optical response of the structure is calculated with sufficient accuracy by considering a cutoff of $N=5$ in the Fourier series expansions involved in our time Floquet scattering-matrix method outlined in section~\ref{sc:theory}.

\begin{figure}
\centering
\includegraphics[width=\linewidth]{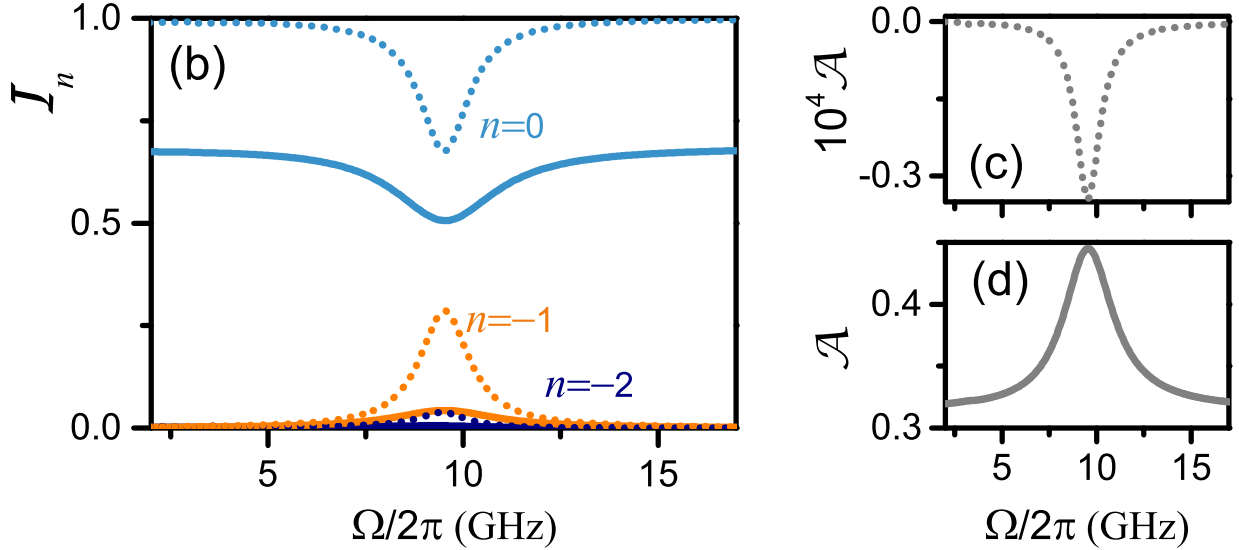}
\caption{The structure of~figure~\ref{fig2}(a), under continuous excitation of a uniform precession spin-wave mode of angular frequency $\Omega$ with a relative amplitude $\eta=0.06$, is illuminated from the left by TM-polarized light with $q_y=3~\mu \mathrm{m}^{-1}$ at the corresponding resonance frequency [see~figure~\ref{fig2}(b)]. Variation of the dominant elastic and inelastic total outgoing light intensities versus the spin-wave frequency (a) and corresponding optical absorption (b) and (c). Dotted and solid curves refer to the lossless and lossy structure, respectively.}
\label{fig4}
\end{figure}
\begin{figure}
\centering
\includegraphics[width=\linewidth]{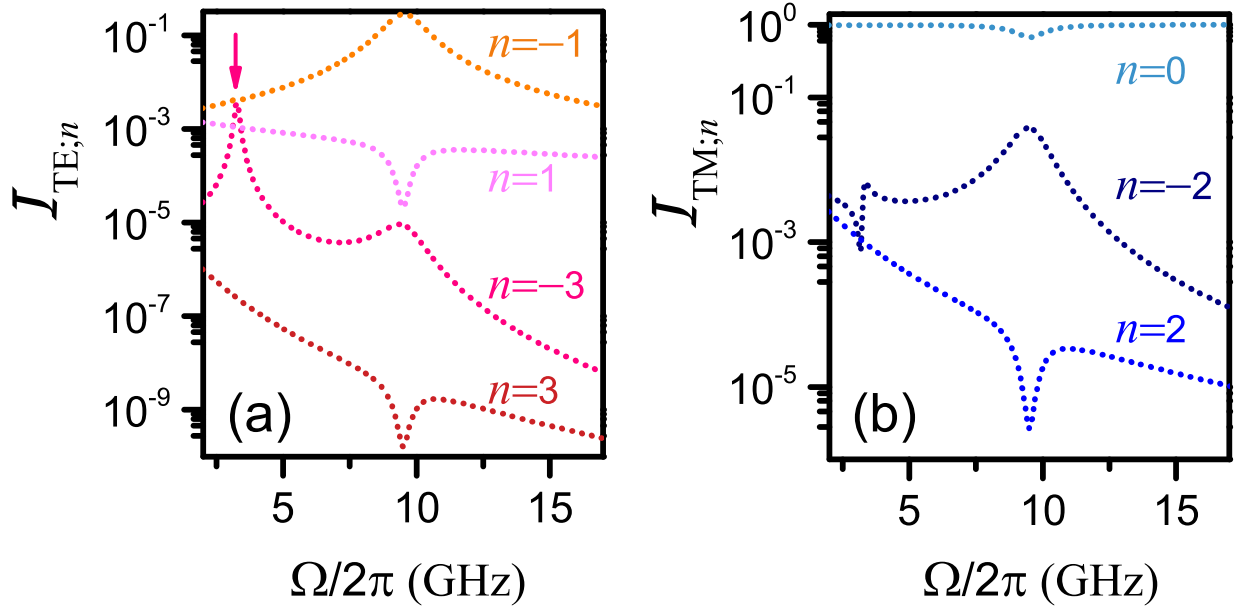}
\caption{Polarization-converting (a) and polarization-conserving (b) contributions to the spectrum of the figure~\ref{fig4}(a). The peak in (a) indicated by the arrow corresponds to the resonant transition when accomplished by absorption of three magnons.}
\label{fig5}
\end{figure}


figure~\ref{fig4}(a) shows the total (transmitted plus reflected) intensities, $\mathcal{I}_n=\sum_{p}(\mathcal{T}_{pn}+\mathcal{R}_{pn})$, as a function of the spin-wave frequency $\Omega/2\pi$.
It can be seen that inelastic light scattering is negligible when the allowed final photon states fall within a gap, where the optical density of states is very low, and we essentially have only the elastic outgoing beam.
On the contrary, when the spin-wave frequency matches the frequency difference $\Delta f=9.5$~GHz between the two optical resonances [see figure~\ref{fig2}(a)], the triple-resonance condition is fulfilled and one-magnon absorption processes are favoured, leading to enhanced intensities of the corresponding ($n=-1$) inelastically transmitted and reflected light beams, with conversion efficiency of the order of 30\% if dissipative losses are neglected.
At the same time, the elastic beam intensity is considerably reduced while the other inelastic processes are also resonantly affected, though to a much lesser degree, as shown in figure~\ref{fig4}(a) and also in figure~\ref{fig5}.

Overall, there is an excess number of magnons absorbed, which can be accounted for by our fully dynamic time Floquet scattering-matrix method~\cite{pantazopoulos2019B}.
This is manifested as a small negative absorption peak [see figure~\ref{fig4}(b)], which clearly indicates a resonant energy transfer from the magnon to the photon field. 

Considering a saturation magnetization $M_0=150~\mathrm{emu/cm^3}$ for Ce:YIG~\cite{onbasli2016}, the triple-resonance condition ($\Omega/2\pi=9.5$~GHz) is achieved with a bias magnetic field $H_0=2.5$~kOe.
In this case, the cone angle of magnetization precession (elliptical in the chosen configuration) attains a maximum of $2.75^\mathrm{o}$, which is a tolerable value for linear spin waves.

It is interesting to note that the triple-resonance condition can be accomplished by many-magnon absorption processes as well ($m\Omega/2\pi=\Delta f$), provided that the number of magnons, $m$, is odd in order to change the polarization state of the photon, from TM to TE, as required in our case.
We recall that our method of calculation is not restricted to the first-order Born approximation and thus it can describe nonlinear effects that are usually relatively weak.
For example, such a three-magnon absorption process is manifested as a peak in the intensity of the $n=-3$ outgoing beam, for $\Omega/2\pi=\Delta f /3\approx3.2$~GHz, as pointed out by the arrow in figure~\ref{fig5}(a). 

As can be seen in figure~\ref{fig4}(a), when dissipative losses are taken into account, the elastic beam intensity is uniformly by about 30\%, in agreement with the results shown in figure~\ref{fig2}(b) for the TM mode.
Here, when the triple-resonance condition is satisfied, the corresponding drop in the $n=-1$ beam is considerably larger because of the longer lifetime of the final (TE) state but, nonetheless, the optical conversion efficiency is still as high as 5\%. 
We note that, because of the high quality factor of the final (TE) state and the presence of non-negligible losses in this case, we overall obtain resonant optical absorption (instead of gain in the lossless case), as shown in figure~\ref{fig4}(c).

\section{Conclusions}
To conclude, we have presented a detailed analysis and optimization of a planar optomagnonic structure operating in the triple-resonance regime and allowing for optical conversion efficiencies of the order of 5\% [cf.~figure~\ref{fig4}(a)] under realistic conditions, mediated by a uniformly precessing spin wave. 
The outlined time Floquet multiple-scattering methodology was able to resolve absorption and emission of multiple magnons, indicating that under special conditions the attained conversion efficiencies mediated by multiple magnons can be comparable to those mediated by a single magnon [cf. orange and pink dotted lines in~figure~\ref{fig5}(a)].
We have also found that the absorption or emission of a magnon leads to a change in the polarization of the optical conversion process.
An interesting further objective would be to extend the current approach to the full \textit{spatio}-temporal Floquet scattering-matrix methodology, which should allow for investigating, among others, surface Dammon-Eshbach and backward volume waves with an in-plane propagation wave vector that can lead to more exotic physical behavior, including emergence of a paraxial outgoing scattered beam and bandgap formation.

\acknowledgments
P.A.P. was supported by
the General Secretariat for Research and Technology (GSRT) and the
Hellenic Foundation for Research and Innovation (HFRI) through a PhD
scholarship (No.~906).
K.L.T., E.A., and G.P.Z. were supported by HFRI and GSRT under Grant 1819.

\end{document}